\documentclass[prd,12pt,draft]{revtex4}

\usepackage{amsmath,amssymb} 
\usepackage[dvips]{graphicx} 
\usepackage[cp1251]{inputenc}

\newcommand\nn{\nonumber}
\newcommand\ba{\begin{eqnarray}}
\newcommand\ea{\end{eqnarray}}
\newcommand{\br}[1]{\left( #1 \right)}
\newcommand{\brs}[1]{\left[ #1 \right]}

\newcommand{\brw}[1]{\left| #1 \right>}

\newcommand{\GeV}{~\mbox{GeV}}
\newcommand{\MeV}{~\mbox{MeV}}
\newcommand{\KeV}{~\mbox{KeV}}
\newcommand{\eV} {~\mbox{eV}}

\newcommand{\nb}{~\mbox{nb}}

\begin{document}

\title{Two photon decay and photoproduction of radial excitation of pion $\pi_0'$}

\author{Eduard~A.~Kuraev}
\email{kuraev@theor.jinr.ru}
\affiliation{JINR-BLTP, 141980 Dubna, Moscow region, Russian Federation}
\author{Mikhail~K.~Volkov}
\email{volkov@theor.jinr.ru}
\affiliation{JINR-BLTP, 141980 Dubna, Moscow region, Russian Federation}

\begin{abstract}
Within the framework of non-local quark model of Nambu--Jona-Lasinio type
the two-photon decay of radial excited state of $\pi_0$-meson -- $\pi_0'$
-- is found to be $3.6\KeV$. The radial excitation of pion is described with the use of
polynomial formfactor of second order over $\vec q^2$ where $\vec q$ is
the transverse momentum, which
corresponds to relative motion of quark-antiquark pair within the energy
range from 0 up to $1\GeV$.
The probabilities of production of $\pi_0'$ and $(\pi_0'+\gamma)$ states in
the electron-positron colliders are estimated. The production of $\pi_0'$-meson
in the interaction of photon with electron or muon is as well considered (Primakoff effect).
The relevant total cross sections are $0.14$ or $0.06\nb$.
\end{abstract}

\maketitle

\section{Introduction}
\label{Introduction}

In recent papers \cite{Bystritskiy:2007wq,Volkov:2008ye,Bartos:2009mx,Volkov:2009mz,Volkov:2009pc}
within the local Nambu--Jona-Lasinio (NJL) model different radiative
decays with the scalar, pseudoscalar and vector mesons of ground state nonets were
calculated.
Some predictions for probabilities of production of meson pairs
$\eta$, $\eta'$ and $f_0(980)$ with $\omega$, $\rho$ and $\phi$
in the collisions in the electron-positron colliders also was done.

During last years the activity in building of the colliders
with high energies of electron-positron beams sufficiently increases
(i.e. BES-III (Beijin, China), CMD-2 (Novosibirsk, Russia), Frascati (Italy)
and others). Thus it appears to be possible to use them for production not
only the ground state of chiral nonets of mesons, but also the
radial excitations of them with the masses from $1.3\GeV$ up to $2\GeV$.
The present paper is devoted to the investigation of radial excitation of pion
$\pi_0'$ production in electron-positron collisions
and to description of two-photon decay of this meson.

For description of radial excitations of mesons the non-local version of
quark NJL model can be used.
In this model the polynomial type of formfactor was proposed. In particular,
for first radial excitation of mesons it is enough to use the simplest
formfactor with square dependence on transverse momentum $\vec k$:
$f(\vec{k})=c(1-d\vec{k}^2)\theta(\Lambda^2-\vec{k}^2)$, where
$\Lambda=1.03\GeV$ is the cut off parameter,
$\vec k$ is the relative quark-antiquark pair momentum, which
takes the values in the range from 0 up to $\Lambda$.
The coefficient $d$ is found from the requirement that scalar excitations do not
give any additional contribution to the quark condensate \cite{Volkov:1996br}.

In papers
\cite{Volkov:1996br,Volkov:1996fk,Volkov:1997dd,Volkov:1999qb,Volkov:1999iq,Volkov:1999xf,Volkov:1999yi,Volkov:2006vq}
this non-local NJL model was successfully applied for description of mass spectrum
of radial excited states of scalar, pseudoscalar and vector mesons.
Besides all the basic strong decays of these mesons was considered and the
results are in satisfactory agreement with the experimental data. In the present
paper we will use this non-local NJL model to calculate the width of two-photon
decay $\pi_0'\to2\gamma$ and to evaluate the probability of production of
$\pi_0'$ meson in the processes $e^+e^- \to e^+e^-\pi_0'$ and
$e^+e^- \to \pi_0'\gamma$. And finally we will consider the so called Primakoff effect
when $\pi_0'$ meson produces in photon-muon scattering.

\section{Quark-meson Lagrangian for ground and radial excited states of the mesons}
\label{Calculation}

The part of lagrangian which we need here is \cite{Volkov:1996br,Volkov:1997dd}:
\ba
\Delta L &=& {\bar q}(p_1)\brs{e Q \gamma_\mu A^\mu(q) +
g_{\pi_1} Z \gamma_5 \pi_1(q) + g_{\pi_2} \gamma_5 \pi_2(q) f\br{\vec q^2}}q(p_2),
\qquad q = p_1-p_2,
\ea
where $e$ is the electron charge ($e^2/4\pi = 1/137$), $Q=diag\br{\frac{2}{3},-\frac{1}{3},-\frac{1}{3}}$
is the quark charge matrix, $A^\mu$ is the electromagnetic field 4-potential,
$\pi_{1,2}$ are non-physical pion fields which interacts with the quarks with the
following coupling constants:
\ba
g_{\pi_1} = \br{4 I_2}^{-1/2}, \qquad
g_{\pi_2} = \br{4 I_2^{ff}}^{-1/2}, \qquad
Z = \br{1-\frac{6m_u^2}{M_{a_1}^2}}^{-1/2} \approx 1.2, \nn
\ea
where $m_u=280\MeV$ is the constituent $u$-quark mass,
$M_{a_1}=1.23\GeV$ is the mass of axial-vector meson $a_1$ \cite{Amsler:2008zzb} and
$I_2$ and $I_2^{ff}$ are the following integrals:
\ba
    I_2 &=& -i\frac{N_c}{\br{2\pi}^4}
    \int d^4 k
    \frac{\theta\br{\Lambda^2-k_\bot^2}}
    {\br{k^2-m^2+i0}^2}, \qquad N_c = 3, \\
    I_2^{ff} &=& -i\frac{N_c}{\br{2\pi}^4}
    \int d^4 k
    \frac{f^2(k_\bot^2)\theta\br{\Lambda^2-k_\bot^2}}
    {\br{k^2-m^2+i0}^2},
    \qquad
    k_\bot = k - \frac{(kq)}{q^2},\nn
\ea
where $k_\bot$ is the momentum transverse to $q$ and in the
meson rest frame ($q=(q_0,0)$): $k_\bot=(0,\vec k)$.
Function $f(k_\bot^2)$ is the form-factor:
\ba
f(k_\bot^2)=c(1-d\vec{k}^2)\theta(\Lambda^2-\vec{k}^2),
\ea
with the restriction imposed on the 3-dimensional momentum of quark interaction.
The satisfactory description leads to the choice of parameters $\Lambda=1.03\GeV$,
$d=1.78\GeV^{-2}$ \cite{Volkov:1996br}.

It is worth notice that states $\brw{\pi_1}$ and $\brw{\pi_2}$ are not
the physical ones since corresponding free Lagrangian contains non-diagonal kinetic terms.
In order to obtain the physical states it is necessary to make an orthogonal transformation \cite{Volkov:1996br}
which will simultaneously diagonalize the mass and $p^2$ terms
in the initial Lagrangian and as a result the wave functions of the
ordinary $\brw{\pi}$ and excited $\brw{\pi'}$ neutral pions are expressed in form of linear
combination of the states $\brw{\pi_1}$ and $\brw{\pi_2}$:
\ba
\brw{\pi_1}&=&\frac{1}{\sqrt{Z_1} \sin\br{2\alpha_0}}
\brs{
    \sin\br{\alpha+\alpha_0} \brw{\pi}-\cos\br{\alpha+\alpha_0}\brw{\pi'}
}, \\
\brw{\pi_2}&=&\frac{1}{\sqrt{Z_2} \sin\br{2\alpha_0}}
\brs{
    \sin\br{\alpha-\alpha_0} \brw{\pi}-\cos\br{\alpha-\alpha_0}\brw{\pi'}
},
\ea
where
$\alpha=59.5^o$ and $\alpha_0=59.15^o$ are the mixing angles evaluated in \cite{Volkov:1997dd}.
%
Further all the interactions of these mesons will be described through the quark loops.

\subsection{Two-photon decay on ground state and radially excited pions}

Consider now the two-photon decays $\pi_0\to 2 \gamma$ and $\pi_0'\to 2 \gamma$.
Matrix elements of the processes $\pi_{1,2}\to 2\gamma$ are
\ba
\frac{3\alpha}{32\pi}\frac{1}{\sqrt{Z_1}}I^1_\Lambda; \qquad
\frac{3\alpha}{32\pi}\frac{1}{\sqrt{Z_2}}I^f_\Lambda,
\ea
with
\ba
I^1_\Lambda=\int\limits_0^A\frac{x^2 dx}{r^5}; \qquad
I^f_\Lambda=\int\limits_0^A\frac{x^2(1-dx^2) dx}{r^5}, \qquad r=\sqrt{x^2+1},
\ea
with $A=\Lambda/m_u=3.68$, $d=Dm_u^2=0.14$ (rest frame of the exited
pion is implied).
Calculation leads to
\ba
I^1_\Lambda=0.297; \qquad I^f_\Lambda=0.193.
\ea
Numerical values of normalization constants are$\sqrt{Z_1}=0.335$; $\sqrt{Z_2}=0.34$.

It is interesting to calculate the width of the ordinary pion in this model.
The result of calculation is
\ba
\Gamma_\pi &=& \br{\frac{\alpha}{\pi}}^2 \frac{1}{64 m_u^2} \frac{M_\pi^3}{\pi}
\brs{
    3 I^1_\Lambda \frac{\sin\br{\alpha+\alpha_0}}{\sqrt{Z_1}\sin\br{2\alpha_0}}
    +
    3 I^f_\Lambda \frac{\sin\br{\alpha-\alpha_0}}{\sqrt{Z_1}\sin\br{2\alpha_0}}
}^2 = 6.5\eV.
\label{PiDecay}
\ea
%
This result are in a good agreement with the experimental
data
\cite{Amsler:2008zzb}: $\Gamma^{exp.}_\pi = 7.8 \pm 0.6 \eV$.
This agreement improves in the limit $\Lambda\to\infty$:
$\Gamma_\pi\br{\Lambda\to\infty} = 7.7\eV$.

Let us note that in strong decays description \cite{Volkov:1997dd}
the similar check was performed for decays $\pi' \to \rho\pi$.
Besides for additional test of this non-local NJL model the decay
$\rho \to 2\gamma$ was considered, which appeared to be also in a
good agreement with the experimental data.

For the exited pion decay width we have
\ba
\Gamma_{\pi'\to2\gamma} &=& \br{\frac{\alpha}{\pi}}^2 \frac{1}{64 m_u^2} \frac{M_{\pi'}^3}{\pi}
\brs{
    3 I^1_\Lambda \frac{\cos\br{\alpha+\alpha_0}}{\sqrt{Z_1}\sin\br{2\alpha_0}}
    +
    3 I^f_\Lambda \frac{\cos\br{\alpha-\alpha_0}}{\sqrt{Z_1}\sin\br{2\alpha_0}}
}^2 = 3.6\KeV.
\label{PiPrimeDecay}
\ea

\subsection{The production of radially excited pion in peripheral electron-
positron collisions}

Matrix element of production of pseudoscalar meson in peripheral electron-
positron collisions
\ba
e(p_1)+\bar{e}(p_2) \to e(p_1')+\bar{e}(p_2')+\pi'(p), \nn
\ea
\ba
s=2p_1p_2, \qquad q_1^2=(p_1-p_1')^2, \qquad q_2^2=(p_2-p_2')^2, \qquad p^2=M^2,
\qquad p_i^2=p_i^{'2}=m_e^2, \nn
\ea
\ba
s \gg M^2 \sim -q_1^2 \sim -q_2^2 \gg m_e^2, \nn
\ea
have a form
\ba
M^{e\bar{e}\to e\bar{e}\pi'}=\frac{4(4\pi\alpha)}{q_1^2q_2^2}N_1N_2
(p_1p_2q_1q_2)I_\pi,
\ea
Where
\ba
N_1=\frac{1}{s}\bar{u}(p_1')\hat{p}_2 u(p_1), \qquad N_2=\frac{1}{s}\bar{v}(p_2)\hat{p}_1 v(p_2'); \nn
\ea
\ba
\sum|N_1|^2=\sum|N_2|^2=2; \qquad
(abcd)=\epsilon_{\alpha\beta\gamma\delta}a^\alpha b^\beta c^\gamma d^\delta.
\ea
The quantity $I_\pi$ can be expressed in terms of two photon decay width $\Gamma^{\pi'\to2\gamma}$
of meson $\pi_0'$:
\ba
|I_\pi|^2=\frac{64\pi\Gamma^{\pi'\to2\gamma}}{M^3}.
\ea
Using the Sudakov parametrization of 4-vectors \cite{Baier:1980kx}:
\ba
q_1&=&\alpha_1\tilde{p}_2+\beta_1 \tilde{p}_1+q_{1\bot}; \nn \\
q_2&=&\alpha_2 \tilde{p}_2+\beta_2 \tilde{p}_1+q_{2\bot},
\ea
with $\tilde{p}_{1,2}$-are the light-like 4-vectors and the on mass shell conditions
of the scattered fermions one can arrange the phase volume of the final state as
\ba
d\Gamma&=&\frac{1}{(2\pi)^5}\frac{d^3 p_1'}{2E_1}\frac{d^3 p_2'}{2E_2}\frac{d^3 p}{2E}\delta^4(p_1+p_2-p_1'-p_2'-p)= \nn \\
&=&(2\pi)^{-5}d^4q_1d^4q_2d^4p_1'd^4p_2'd^4p\delta^4(q_1+q_2-p)\delta^4(p_1-q_1-p_1')\delta^4(p_2-q_2-p_2')
\times\nn\\
&\times&\delta(p^2-M^2)\delta(p_1^{'2}-m_e^2)\delta(p_2^{'2}-m_e^2) =\nn\\
&=&(2\pi)^{-5}\frac{d\beta_1}{4s\beta_1}d^2 \vec{q}_1d^2\vec{q}_2,
\ea
where we imply
\ba
\frac{M^2}{s}<\alpha_2\sim \beta_1<1; \qquad s\alpha_2\beta_1=M^2+(\vec{q}_1+\vec{q}_2)^2; \nn \\
q_{1\bot}^2=-\vec{q}_1^2, \qquad q_{2\bot}^2=-\vec{q}_2^2, \vec{q}_i^2\sim M^2. \nn
\ea
The transferred momentum squares are negative quantities
\ba
q_1^2\approx-(\vec{q}_1^2+m_e^2\beta_1^2); \qquad
q_2^2\approx-(\vec{q}_2^2+m_e^2\alpha_2^2).
\ea
The differential cross section have a form:
\ba
d\sigma^{e\bar{e}\to e\bar{e}(\pi,\pi')}&=&
8\tilde\sigma_{(\pi,\pi')} \frac{d\beta_1}{\beta_1}
\frac{d\vec{q}_1^2 d\vec{q}_2^2 \vec{q}_1^2 \vec{q}_2^2}{(\vec{q}_1^2+\beta_1^2m_e^2)^2
(\vec{q}_2^2+m_e^2(\frac{M^2}{s\beta_1})^2)^2}, \\
\tilde\sigma_\pi &=& \frac{\alpha^2 \Gamma^{\pi\to2\gamma}}{M_\pi^3} = 0.8 \cdot 10^{-4}\nb, \nn\\
\tilde\sigma_{\pi'} &=& \frac{\alpha^2 \Gamma^{\pi'\to2\gamma}}{M_{\pi'}^3} = 3.3 \cdot 10^{-5}\nb. \nn
\ea
For the case of light mesons ($\pi$) production we obtain, performing the integration:
\ba
\sigma_\pi=8\tilde\sigma_\pi J(s),
\ea
\ba
J(s)=L(l+L-1)^2-\frac{13}{3}L^3, \qquad L=\ln\frac{s}{M_\pi^2}, \qquad l=\ln\frac{M_\pi^2}{m_e^2}. \nn
\ea
For the case of heavy meson production we have \cite{Brodsky:1970vk}
\ba
\sigma_{\pi'}=2\tilde\sigma_{\pi'}Y\br{\frac{M_{\pi'}^2}{s}} l^2;
\qquad
Y(z)=(2+z)^2\ln\frac{1}{z}-(3+z)(1-z).
\ea
If $\sqrt{s} = 3\GeV$ then $\sigma_\pi = 0.43\nb$ and $\sigma_{\pi'} = 0.09\nb$.

\subsection{The annihilation channel of radiative production of excited neutral pion}

Let consider the annihilation channel of process radiative production of excited neutral
pion in electron-positron annihilation:
\ba
e^-(p_-)+e^+(p_+) \to \gamma^* \to \pi'(p)+\gamma(k), \qquad q=p^++p_-.
\ea
 The corresponding matrix element is
 \ba
M=\frac{(4\pi\alpha)^{1/2}}{s}\bar{v}(p_+)\gamma_\mu u(p_-)\frac{2 m_u \alpha}{\pi s}\epsilon_{\mu\nu\lambda\sigma}~R~
e^\nu q^\lambda k^\sigma,
\label{EEPiGammaAmplitude}
 \ea
where $e_\nu(k)$ is the polarization 4-vector of the real photon.
The quantity $R=-0.302$ will be calculated in Appendix~\ref{Appendix}.
Expressing the phase volume of the final state as
\ba
d\Gamma_2&=&(2\pi)^{-2}\frac{d^3p}{2E}\frac{d^3k}{2\omega}\delta^4(P_++p_--p-k)= \nn \\
&=&\frac{1}{16\pi}\br{1-\frac{M^2}{s}}d c, \qquad c=\cos\theta,
\ea
with $\theta$ is the angle of exited pion emission regarding the beams axis ($\vec{p}_-$)
in the beams center of mass frame.
After simple manipulations the expression for the differential cross section will be
(see (\ref{A7}))
\ba
\frac{d\sigma^{e\bar{e}\to \pi'\gamma}}{d c}=\frac{\alpha^3m_u^2}{64s^2\pi^2}
\br{1-\frac{M^2}{s}}^3
|R|^2[1+c^2], \qquad R=-0.57.
\ea
If $\sqrt{s}=3\GeV$, then
\ba
    \left.\frac{d\sigma^{e\bar{e}\to \pi'\gamma}}{d c}\right|_{\theta=5^0} = 8.2 \cdot 10^{-8}\nb,
    \qquad
    \left.\frac{d\sigma^{e\bar{e}\to \pi'\gamma}}{d c}\right|_{\theta=45^0} = 1.8 \cdot 10^{-8}\nb.
    \nn
\ea

Keeping in mind application to the real experimental data the possible
intermediate states $\rho'$ and $\omega'$ must be taken into account.

\subsection{Primakoff process}
\label{Primakoff}

Consider now the Primakoff process with creation of the excited pion state
in lepton-photon high energy collisions:
\ba
\gamma(k)+l(p)\to \pi'(p_1)+l(p'),
\qquad
l = e,\mu,
\ea
\ba
p^2=p^{'2}=m_l^2, \qquad k^2=0, \qquad p_1^2=M^2, \qquad s=2kp > M^2 \gg m_l^2.
\nn
\ea
Matrix element have a form
\ba
M=\frac{\sqrt{4\pi\alpha}}{q^2}\bar{u}(p')\gamma_\mu u(p) I_\pi(\mu\epsilon k q),
\ea
with $\epsilon$ is the polarization vector of the photon.
Using the Sudakov parametrization of the transferred 4-momentum $q=p-p'$:
\ba
q=a\tilde{p}+\beta k+q_\bot, \qquad q^2=-\frac{\vec{q}^2+a^2m_l^2}{1-a}, \qquad a=\frac{\vec{q}^2+M^2}{s},
\ea
we obtain for the cross section
\ba
d\sigma^{\gamma e\to \pi' e}&=&\frac{\alpha\Gamma}{M^3}\br{1+\br{1-\frac{M^2}{s}}^2}
\frac{\vec{q}^2 d\vec{q}^2}{\br{\vec{q}^2+m_l^2\frac{M^4}{s^2}}^2}; \nn \\
\sigma^{\gamma e\to \pi' e}&=&\frac{\alpha\Gamma}{M^3}\br{1+\br{1-\frac{M^2}{s}}^2}
\br{\ln\frac{s^2}{m_l^2M^2}-1}.
\ea
The total cross section for $\sqrt{s}=3\GeV$ is $\sigma^{\gamma e\to \pi' e}=0.14\nb$
and $\sigma^{\gamma \mu\to \pi' \mu}=0.06\nb$.

\section{Conclusion}
\label{Conclusion}

In our calculations besides the prediction for the width of
two-photon decay of $\pi_0'$ meson we also obtained the two-photon decay
width of ground state $\pi_0$ meson, which is in a good agreement with the
corresponding experimental data. The similar result appears in the
standard local NJL model \cite{Volkov:1986zb}.
This result confirms the statement of \cite{Volkov:1999yi} that
the introduction of radial excitations of mesons into non-local NJL model
does not distort the picture which appeared in local NJL model in
description of ground state mesons.
The same conclusion was done during the consideration of
strong decays within the same non-local NJL model \cite{Volkov:1997dd,Volkov:1999yi}.

We should note that in photoproduction of $\pi_0'$ meson in electron-positron
colliders in the final state two charged mesons $\rho^\pm$ and $\pi^\mp$
(which the $\pi_0'$ meson will decay to) will be
detected. But in case if there is no $\pi_0'$ meson in the intermediate state,
but the annihilation results in simple virtual photon $\gamma^*$, only
the neutral pair $(\rho_0,\pi_0)$ will be produced.

In future we're planning to perform the similar analysis for
radial excitations of $\eta$ and $\eta'$ mesons.

\begin{acknowledgments}
The authors thank A.~B.~Arbuzov for critical comments and independent derivation of
Eqs. (\ref{PiDecay}) and (\ref{PiPrimeDecay}).
The authors wish to thank Yu.~M.~Bystritskiy for help in paper preparation.
We also acknowledges the support
of INTAS grant no. 05-1000008-8528 and 
RFBR grant 10-02-01295-a.
\end{acknowledgments}

\appendix

\section{The amplitude $\gamma^*\to\gamma\pi'$}
\label{Appendix}

Consider now the amplitude of conversion of a heavy photon to the real one and the excited pion
\ba
\gamma^*(q) \to \gamma(k)+\pi'(p), \qquad k^2=0; \qquad q=\sqrt{s}(1,0,0,0),
\ea
where we suppose photon energy $\omega=(s-M^2)/(2\sqrt{s})$ is much smaller than the mass of the
excited pion and we can consider the exited pion almost at rest.

We must estimate the loop integrals with the 3-dimensional cut-off:
\ba
J_\Lambda^{1,f}&=&\int\frac{d^4 k m}{(2\pi)^4}\frac{\Theta(\Lambda-|\vec{k}|)[1;f(\vec{k}^2)]}{d_0d_1d_2}, \nn \\
d_0\approx d_1&=&k^2-m^2+i0=(k_0-\epsilon_k+i0)(k_0+\epsilon_k-i0); \nn \\
d_2&=&(q+k)^2-m^2+i0=[k_0+\sqrt{s}-\epsilon_k+i0][k_0+\sqrt{s}+\epsilon_k-i0].
\ea
Performing the $k_0$ integration and using the dimension-less variable $x=|\vec{k}|/m$, we obtain:
\ba
J_\Lambda^{1,f}=\frac{m}{8\pi^2 s}I^{1,f},
\ea
with
\ba
I^{1,f}&=&\int\limits_0^A\frac{x^2 dx[1,f]}{r^3}\frac{1}{1-2\eta}[1-\frac{2\eta(1-\eta)}{1-2\eta}-
\frac{2\eta r}{(1-2\eta)^2}], \\
\eta&=&\frac{m^2 r}{s}, \qquad r=\sqrt{x^2+1}. \nn
\ea
At small values of $\eta$ we have
\ba
I^{1,f} \to I_{as}^{1,f}=\int\limits_0^A\frac{x^2 dx[1,f]}{r^3},
\ea
\ba
f=1-0.14x^2;  \qquad I_{as}^1=1.05; \qquad I_{as}^f=0.35.
\ea
The quantity $R$ entering (\ref{EEPiGammaAmplitude}) is
\ba
R=
\frac{\cos\br{\alpha+\alpha_0}}{\sin\br{2\alpha_0} \sqrt{Z_1}} I_{as}^1
+
\frac{\cos\br{\alpha-\alpha_0}}{\sin\br{2\alpha_0} \sqrt{Z_2}} I_{as}^f
=
-0.57.
\label{A7}
\ea


\begin{thebibliography}{10}

\bibitem{Bystritskiy:2007wq}
Y.~M. Bystritskiy, M.~K. Volkov, E.~A. Kuraev, E.~Bartos, and M.~Secansky,
\newblock Phys. Rev. {\bf D77}, 054008 (2008), 0712.0304.

\bibitem{Volkov:2008ye}
M.~K. Volkov, Y.~M. Bystritskiy, and E.~A. Kuraev,
\newblock (2008), 0811.3773.

\bibitem{Bartos:2009mx}
E.~Bartos, M.~Secansky, Y.~M. Bystritskiy, E.~A. Kuraev, and M.~K. Volkov,
\newblock (2009), 0902.1384.

\bibitem{Volkov:2009mz}
M.~K. Volkov, Y.~M. Bystritskiy, and E.~A. Kuraev,
\newblock (2009), 0901.1981.

\bibitem{Volkov:2009pc}
M.~K. Volkov, E.~A. Kuraev, and Y.~M. Bystritskiy,
\newblock (2009), 0904.2484.

\bibitem{Volkov:1996br}
M.~K. Volkov and C.~Weiss,
\newblock Phys. Rev. {\bf D56}, 221 (1997), hep-ph/9608347.

\bibitem{Volkov:1996fk}
M.~K. Volkov,
\newblock Phys. Atom. Nucl. {\bf 60}, 1920 (1997), hep-ph/9612456.

\bibitem{Volkov:1997dd}
M.~K. Volkov, D.~Ebert, and M.~Nagy,
\newblock Int. J. Mod. Phys. {\bf A13}, 5443 (1998), hep-ph/9705334.

\bibitem{Volkov:1999qb}
M.~K. Volkov, V.~L. Yudichev, and M.~Nagy,
\newblock Nuovo Cim. {\bf A112}, 955 (1999).

\bibitem{Volkov:1999iq}
M.~K. Volkov and V.~L. Yudichev,
\newblock Int. J. Mod. Phys. {\bf A14}, 4621 (1999), hep-ph/9904226.

\bibitem{Volkov:1999xf}
M.~K. Volkov and V.~L. Yudichev,
\newblock Phys. Atom. Nucl. {\bf 63}, 1835 (2000), hep-ph/9905368.

\bibitem{Volkov:1999yi}
M.~K. Volkov and V.~L. Yudichev,
\newblock Phys. Part. Nucl. {\bf 31}, 282 (2000), hep-ph/9906371.

\bibitem{Volkov:2006vq}
M.~K.~Volkov and A.~E.~Radzhabov,
\newblock Phys.\ Usp.\  {\bf 49}, 551 (2006).

\bibitem{Amsler:2008zzb}
C.~Amsler {\it et al.}  [Particle Data Group],
\newblock Phys.\ Lett.\  B {\bf 667}, 1 (2008).

\bibitem{Baier:1980kx}
V.~N. Baier, E.~A. Kuraev, V.~S. Fadin, and V.~A. Khoze,
\newblock Phys. Rept. {\bf 78}, 293 (1981).

\bibitem{Brodsky:1970vk}
S.~J. Brodsky, T.~Kinoshita, and H.~Terazawa,
\newblock Phys. Rev. Lett. {\bf 25}, 972 (1970).

\bibitem{Volkov:1986zb}
M.~K. Volkov,
\newblock Sov. J. Part. and Nucleai {\bf 17}, 186-203 (1986).

\end{thebibliography}

\end{document}